\documentclass[%
 reprint,
superscriptaddress,
 amsmath,amssymb,
 aps,
pra,
showkeys
]{revtex4-2}

\usepackage{graphicx}
\usepackage{dcolumn}
\usepackage{bm}
\usepackage{float}
\usepackage{graphicx}
\usepackage{dcolumn}
\usepackage{bm}
\usepackage{upgreek}
\usepackage{xcolor} 
\usepackage{soul,xcolor}
\usepackage{amsmath}
\usepackage{fancyhdr}
\begin{document}

\newcommand{\Ecomment}[1]{\textcolor{orange}{#1}}
\newcommand{\Estrike}[1]{\setstcolor{orange}{#1}}
\newcommand{\Lucomment}[1]{\textcolor{blue}{#1}}
\newcommand{\Lustrike}[1]{\setstcolor{blue}{#1}}
\newcommand{\krb}[1]{\textcolor{purple}{krb: #1}}

\preprint{APS/123-QED}

\title{Experimental evidence for dipole-phonon quantum logic in a trapped calcium monoxide and calcium ion chain}

\author{Lu Qi} 
\thanks{Contributed equally.}
\email{lu.qi@duke.edu}
\affiliation{Duke Quantum Center, Duke University, Durham, NC 27701, USA}
\affiliation{Department of Electrical and Computer Engineering, Duke University, Durham, NC 27708, USA}
\author{Evan C. Reed}
\thanks{Contributed equally.}
\email{er118@duke.edu}
\altaffiliation[Current Address: ]{IonQ Inc., 4505 Campus Dr, College Park, MD 20740, USA}

\affiliation{Duke Quantum Center, Duke University, Durham, NC 27701, USA}
\affiliation{Department of Electrical and Computer Engineering, Duke University, Durham, NC 27708, USA}

\author{Boyan Yu}
\altaffiliation[Current Address: ]{Department of Physics and Astronomy, University of British Columbia, 6224 Agricultural Rd, Vancouver, B.C., V6T 1Z1, Canada}
\affiliation{Duke Quantum Center, Duke University, Durham, NC 27701, USA}
\affiliation{Department of Electrical and Computer Engineering, Duke University, Durham, NC 27708, USA}

\author{Kenneth R. Brown}
\email{kenneth.r.brown@duke.edu}
\affiliation{Duke Quantum Center, Duke University, Durham, NC 27701, USA}
\affiliation{Department of Electrical and Computer Engineering, Duke University, Durham, NC 27708, USA}
\affiliation{Departments of Physics and Chemistry, Duke University, Durham, NC 27708, USA}

\date{\today}

\begin{abstract}
Dipole-phonon quantum logic (DPQL) offers novel approaches for state preparation, measurement, and control of quantum information in molecular ion qubits. In this work, we demonstrate an experimental implementation of DPQL with a trapped calcium monoxide and calcium ion chain at room temperature. We present evidence for one DPQL signal in two hours of data collection. The signal rises clearly above the characterized noise level and has a lower bound on the statistical significance of 4.1$\sigma$. The rate of observation is limited by the low thermal population in the molecular ground rotational state. 

\end{abstract}

\maketitle


\section{\label{sec:Intro}Introduction}

The coherent control of atoms has become an essential resource for studies of quantum physics and engineering quantum technologies. 
Diatomic molecules, with their additional vibrational and rotational degrees of freedom, provide a broader spectrum that can be utilized for various purposes. 
The coherent control of the rich internal states of molecules has opened new opportunities for precision measurements~\cite{acme2018improved}, studies of quantum chemistry~\cite{doi:10.1126/science.adl6570, karman2024ultracold}, quantum simulation~\cite{li2023tunable,chen2023field} and quantum information processing~\cite{doi:10.1126/science.adf8999, doi:10.1126/science.adf4272}.
Like neutral molecules, the utility of molecular ions has been demonstrated in these applications~\cite{doi:10.1126/science.adg4084, doi:10.1126/science.aan4701,doi:10.1126/science.aba3628,staanum2010rotational,germann2014observation,kortunov2021proton,calvin2018rovibronic,lin2020quantum}.  
In molecular ion research, quantum logic spectroscopy (QLS) is a powerful tool for state preparation and measurement of internal states of single molecular ions~\cite{Chou2017,Wolf2016,doi:10.1126/science.aaz9837,holzapfel2024quantum}. 
Originally developed for use in trapped atomic ion clocks~\cite{Schmidt_QLS_2005}, QLS transfers the quantum information from a trapped molecular ion's internal state to its motional state, which is then read out via a co-trapped atomic ion. 
This approach circumvents the need for molecular transitions with high photon cycling rates, often required for laser cooling in neutral molecule platforms~\cite{shuman2010laser}, and is compatible with various molecular transitions, including hyperfine~\cite{Chou2017,holzapfel2024quantum}, rotational~\cite{doi:10.1126/science.aba3628,doi:10.1126/science.ado1001}, rovibrational, and rovibronic transitions~\cite{Wolf2016,doi:10.1126/science.aaz9837}.
Recently, a novel approach to QLS was proposed that uses the interaction of the permanent dipole of a polar molecular ion with the trap electric field.
In this scheme known as dipole-phonon quantum logic (DPQL), opposite-parity states, such as $\Lambda$ and $\Omega$ doublet states~\cite{Campbell_Hudson_2020} of certain rotational states, are used to facilitate the interaction.
By carefully selecting the appropriate diatomic molecule, a doublet frequency splitting can be found that is comparable to the trapped ions' secular oscillation frequency such that the doublet states may couple to and directly exchange energy with the motional mode~\cite{mills2020dipole}. 
This technique offers a new set of tools for molecular state preparation and measurement and may even be used for generating atom-molecule entanglement as well as for enabling long distance interactions between molecular ions via virtual phonons across a trapped ion chain~\cite{Campbell_Hudson_2020}. 

\begin{figure*}[tp]
    \centering
    \includegraphics[width=0.8\linewidth]{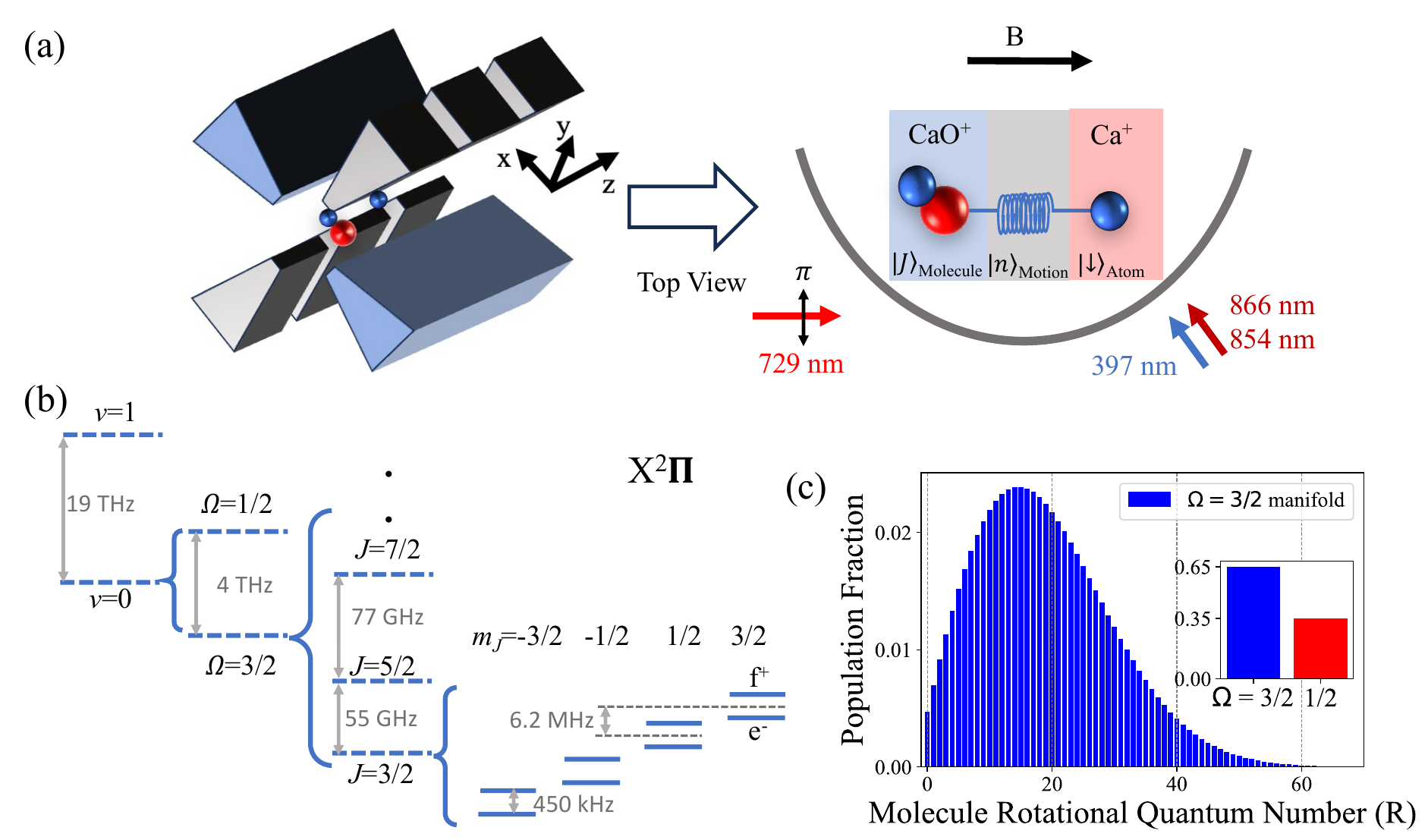}
    \caption{Experimental setup and the energy structure and population distribution of CaO$^+$ under room termperature. 
    (a) A schematic of the experimental setup for the demonstration of dipole-phonon quantum logic. 
    A single CaO$^+$ and Ca$^+$ are trapped in a segmented blade trap. 
    A magnetic field of 6.2\,G is aligned to the trap axis ($z$) to provide the quantization axis and break state degeneracies. 
    A top view of the trap setup shows the laser wavelengths and orientations in which 397\,nm and 866\,nm are used for Doppler cooling, 729\,nm laser for $S_{1/2} \leftrightarrow D_{5/2}$ operation, and 854\,nm laser for quenching $D_{5/2}$ population. 
    (b) The energy level structure of CaO$^+$ ground electronic state X$^2\Pi$ (not to scale) in the presence of a 6.2\,G magnetic field. 
    e$^-$/f$^+$ indicates the odd/even parity state of the $\Omega$-doublets. 
    (c) The distribution of the first 71 rotational states in $\Omega$=3/2, $v=0$. (Inset) The distribution of the two fine structure states, $\Omega$, within each rotational manifold at room temperature. The calculation of the distribution is in Appendix A.
    }
    \label{fig:schematicofexperiment}
\end{figure*}

In this work, we present evidence for the coherent control and detection of DPQL between a calcium oxide ion (CaO\(^+\)) and the trap electric potential. 
In CaO\(^+\), the cation Ca\(^{2+}\), with a closed shell, and the anion O\(^-\), with a half-filled orbital, form an ionic bond. 
The half-filled 2p$\pi$ orbital results in a ground state of \(X^2\Pi_{3/2}\) for CaO\(^+\), while the first excited 2p\(\sigma\) orbital gives rise to a low-lying \(A^2\Sigma^+\) electronically excited state. 
The combination of the rotational electronic Coriolis effect and the spin-orbit coupling between these two molecular orbitals leads to a degeneracy splitting of the \(\Omega\)-doublets, and the magnitude of this splitting increases with rotational number $R$ \cite{mills2020dipole}. 
Specifically, for the ground rotational state with total angular momentum $J=R+\Lambda +\Sigma =3/2$, where $\Lambda=1$ and $\Sigma=1/2$ are the projection of the electron orbital angular momentum and the electron spin on the internuclear axis respectively, the energy splitting between the lower energy doublet state $|e\rangle$ and higher energy doublet state $|f\rangle$ is calculated to be approximately \(\omega_{\text{mol}} = 2\pi \times 450\,\text{kHz}\)~\cite{mills2020dipole} based on previously reported spectroscopic data~\cite{VANGUNDY201817}.

In a linear Paul trap, CaO\(^+\) exhibits secular motion around its equilibrium position such that the ion experiences an oscillating electric field due to this motion. 
When this oscillation frequency $\omega_{q}$ is near resonance with \(\omega_{\text{mol}}\), the interaction between the electric field and the molecular dipole can be described by the Jaynes-Cummings Hamiltonian under the rotating wave approximation~\cite{Campbell_Hudson_2020}:
\begin{equation}
    H_{\text{I}}/\hbar = \frac{\omega_{mol}}{2} \sigma_z + \omega_q \left(a_q^{\dagger} a_q + \frac{1}{2}\right) + \frac{g_q}{2} \left(a_q \sigma_{+} + a_q^{\dagger} \sigma_{-}\right),
\end{equation}
where \(g_q\) is the vacuum Rabi frequency. 
For the ground rotational state, the vacuum Rabi frequency is calculated to be \(2\pi \times 2.6\,\text{kHz}\). 
Because the frequency of this \(\Omega\)-doublet has not yet been directly measured and the dipole-phonon coupling strength is small compared to the doublet splitting, we utilize adiabatic passage with constant ramping speed to facilitate the dipole-phonon interaction. 
By adiabatically ramping the trap frequency $\omega_q$ through resonance with \(\omega_{\text{mol}}\), the \(\Omega\)-doublet states will flip causing a corresponding change in the phonon state of the ion’s motion: $|f,n\rangle \leftrightarrow |e,n+1\rangle$. 
A calcium ion (Ca\(^+\)) is co-trapped with the CaO\(^+\) to provide sympathetic ground state cooling and read out the phonon state. 

\section{\label{sec:exp}Experiment}
A schematic of our experimental system is shown in Fig.~\ref{fig:schematicofexperiment}. Our experiment begins by preparing the CaO\(^+\)-Ca\(^+\) ion chain. 
First, two Ca\(^+\) ions are loaded into a segmented blade trap in an ultra-high vacuum (UHV) chamber~\cite{Qi2023}. 
Subsequently, O\(_2\) gas is introduced via a leak valve while the 397\,nm laser is tuned closer to the \(S_{1/2} \rightarrow P_{1/2}\) resonance of Ca\(^+\) to increase the population in the \(P_{1/2}\) state to facilitate the reaction between Ca$^+$ and O$_2$~\cite{Schmid_2019,mills2020dipole}. 
During this process, the vacuum pressure is maintained at \(1.5 \times 10^{-9}\,\text{Torr}\), and it returns to normal pressure of about \(3 \times 10^{-10}\,\text{Torr}\)  within 5 minutes after the leak valve is closed. 
The vacuum pressure is deduced from a combination of ion pump reading and the ion position exchange rate~\cite{OKA2021111482}.
The leak valve is closed once one of the two trapped Ca$^+$ ions reacts with O$_2$, thereby forming CaO$^+$. 
The axial in-phase (IP) mode secular frequency is checked by exciting the ions' motion with an oscillating trap electrode and observing the fluorescence change ~\cite{10.1116/1.570570}. This is performed to confirm that the mass of the dark ion is 56 amu.

\begin{figure}[h]
    \centering
    \includegraphics[width=\linewidth]{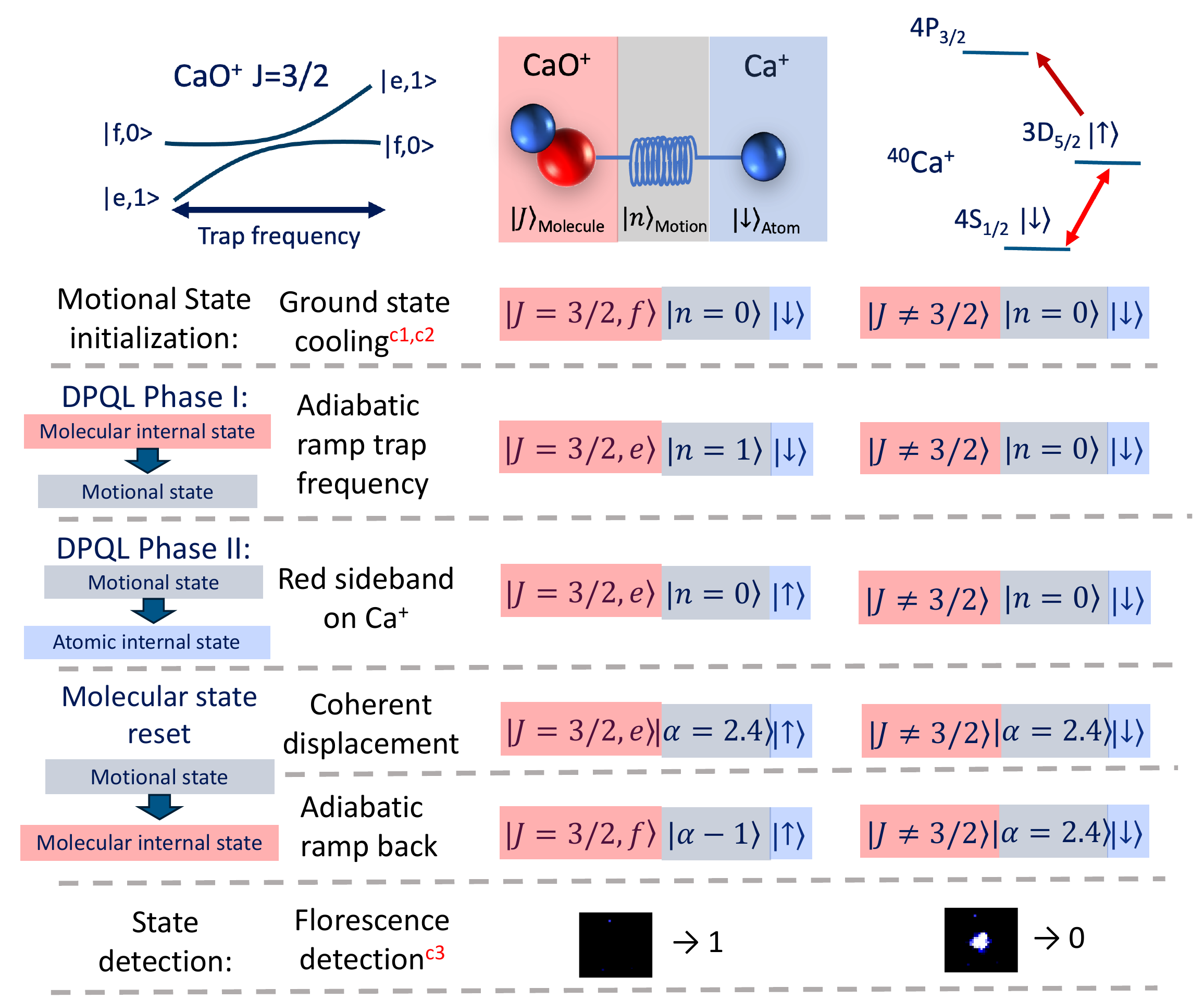}
    \caption{Experimental sequence for DPQL. The avoided crossing on the top left shows the state trajectories in the adiabatic passage operation. The energy structure of Ca$^+$ on the top right illustrates the atomic qubit states used in the experiment. 
    Shown here is every step of the experiment as well as the corresponding state changes of the molecular ion (both $J=3/2$ and $J\neq 3/2$), motion, and atomic ion. The results of single shot measurements of the Ca$^+$ ion are assigned the values of 1 for dark and 0 for bright. c1:\:Dark ion mass check. c2:\:Prior selection check of Ca$^+$ fluorescence. c3:\:Posterior selection check operation. See text for details.
    }
    \label{fig:DPQLsequence}
\end{figure}

Once CaO$^+$ is formed, we wait for 5 minutes for the vacuum pressure to settle and allow the rotational state distribution to thermalize to the ambient blackbody radiation (BBR) of the environment.
During this period of time, we calibrate the laser frequencies and validate the sideband cooling and adiabatic ramping operation~\cite{Qi2023,PhysRevA.110.013123} (Appendix B).
After this period,  we start the search for the DPQL signal following the procedure shown in Fig.~\ref{fig:DPQLsequence}. 
The procedure begins with Doppler cooling the 3-dimensional motion of the ion chain, which is followed by sideband cooling of the IP and out-of-phase (OP) modes of axial motion to the ground motional state.
At this point, before the adiabatic ramping of the trap frequency, the secular frequencies of the IP and OP modes are 2$\pi\times$275\,kHz and 2$\pi\times$492\,kHz, respectively.
After ground state cooling, adiabatic ramping of the OP mode frequency from 2$\pi\times$492\,kHz to 2$\pi\times$410\,kHz is performed with a constant ramping speed of 2$\pi\times$10\,kHz/ms over a duration of 8\,ms to induce the interaction between the $\Omega$-doublet opposite-parity states within the same $m_J$ manifold. 
When the ramping is over, a red sideband pulse on $S_{1/2} \leftrightarrow D_{5/2}$ of the Ca$^+$ ion maps the phonon state to the atomic state. 
If the interaction occurred, then a phonon would have been emitted into the motional mode, and the red sideband on the Ca$^+$ would excite the internal state of the atom to the $D_{5/2}$ state. 
If there was no interaction, then the atom would remain in the $S_{1/2}$ state after the sideband operation.
Before detecting the atomic state, a calibrated RF pulse with frequency 2$\pi\times$410\,kHz is applied to one of the end cap electrodes of the Paul trap to displace the OP mode to a coherent state $|\alpha\rangle$~\cite{Meekhof_Wineland_1996} with $\alpha = 2.4$. 
Following this, the OP mode frequency is adiabatically ramped back to 2$\pi\times$492\,kHz to reset the molecular ion’s internal state.
At the end of the procedure, the atomic state is measured by driving the $S_{1/2} \rightarrow P_{1/2}$ transition and counting the scattered photons. 
In an ideal experiment, a dark result indicates that the atomic ion is in $D_{5/2}$ and the occurrence of the dipole-phonon interaction, whereas a bright result indicates that the atomic ion is in $S_{1/2}$ state and no interaction took place during the adiabatic ramping.

To enhance the reliability of the experimental results, we implement redundant checks to eliminate false positive outcomes from background gas collisions and motional state cooling failures. 
The first check is the mass check of the dark ion. It is applied every 10,000 experiments to confirm the dark ion is CaO$^+$. 
Only data between two validated check results are used for data analysis.
The second check is applied at the beginning of the Doppler cooling stage to confirm the Ca$^+$ is fluorescing at normal level. 
If not, the experiment will halt until the check is passed. 
The third check is a posterior selection check and is applied after atomic state detection. 
This check includes a 854\,nm quenching pulse and a second atomic state detection. 
If a dark result is obtained in the experiment and a bright result is obtained in this check, the dark result is validated and can be used for data analysis. 
If the result of the check is also dark, the corresponding dark experiment result will be discarded.
After these checks, the probability of a dark result, which is a false positive, is 0.03 and is largely due to off-resonant coupling to the carrier transition while shelving the Ca$^+$ with red sideband transition (Appendix B). 
Conversely, when the molecule starts in the DPQL state but does not result in a dark result this is a false negative. The probability of a false negative is a combination of imperfections mainly from motional decoherence in three operations: fidelity of ramping with ground state $p_{r1} = 0.9$ ($|J=3/2, m_J, f\rangle \rightarrow |J=3/2, m_J, e\rangle$), fidelity of ramping with the coherent state $p_{r2} = 0.85$ ($|J=3/2, m_J, e\rangle \rightarrow |J=3/2, m_J, f\rangle$) and shelving fidelity $p_d$ = 0.95. 
This in total gives a fidelity of $p_{r1}\times p_{r2} \times p_d = 0.72$, which increases the probability of a dark result from 0.03 to 0.72 if CaO$^+$ is in ground rotational state.

For CaO\(^+\), the probability of population of the ground rotational state 
at room temperature is only 0.47\% due to its small rotational constants (Appendix A) as shown in Fig.~\ref{fig:schematicofexperiment}(c).
Therefore, the experiment needs to be repeated many times in order to prepare the qubit space of the molecule via projective measurement~\cite{Vogelius_2006}. 
We repeat the experiment 30,000 times per search trial. 
Each experiment takes 38\,ms and a search trial takes 20\,minutes. 
Upon successful projective measurement, the ground state population will start to rethermalize to the BBR with a predicted 1/e time of 4 seconds (Appendix A).
The lifetime of CaO$^+$ in our system is approximately 20\,minutes after which it becomes CaOH$^+$ in most cases. 

\section{\label{sec:dis}Results and Discussions}
\begin{figure}[tp]
    \centering
    \includegraphics[width=\linewidth]{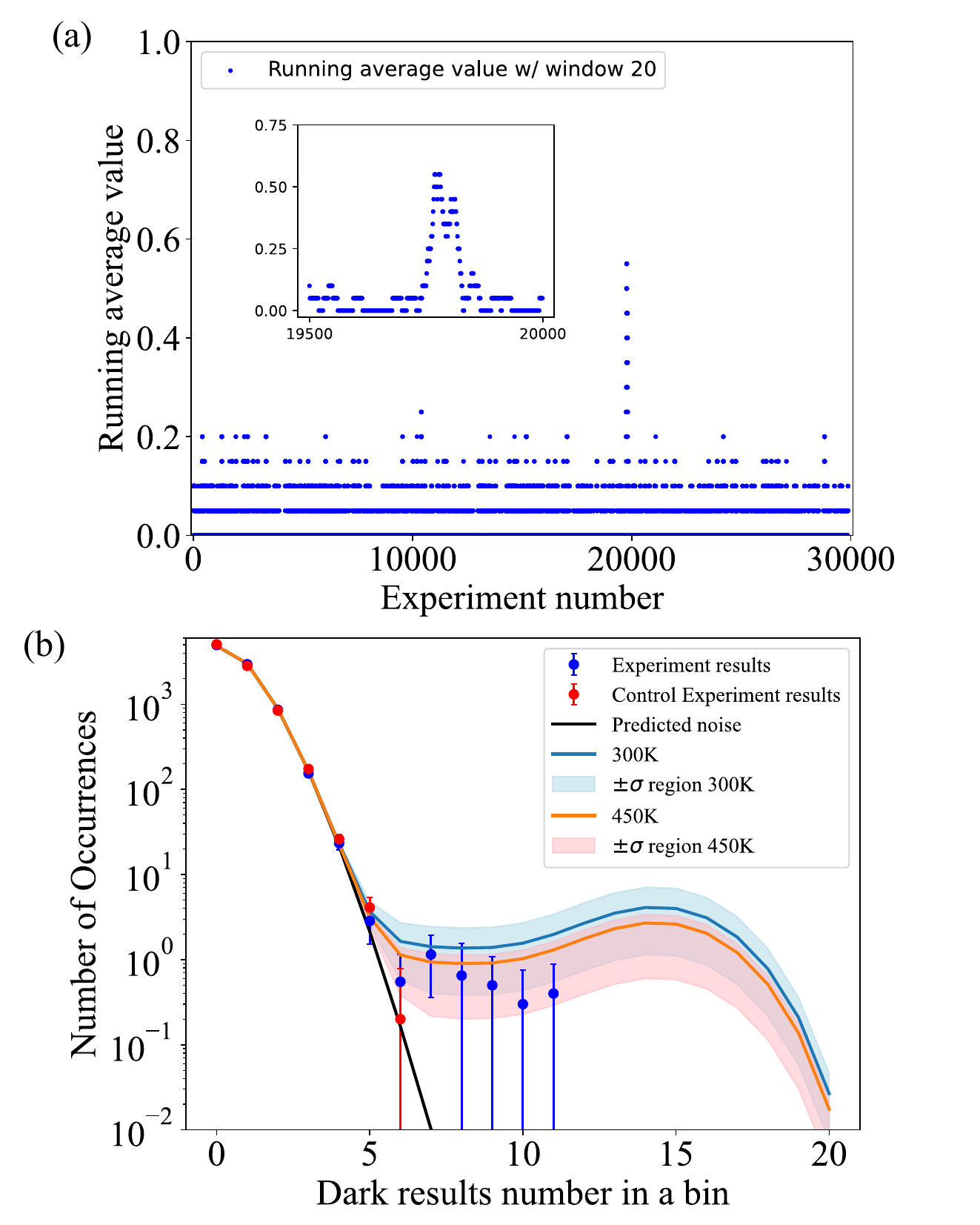}
    \caption{ Experimental results of the DPQL signal search. (a) Experimental results from one of six DPQL search trials in which a suspected DPQL signal is present. Each point represents a moving average of single shot measurements with a bin size of 20 experiments. (Inset) Zoomed-in view of the identified signal. (b) Bin value distributions of 2-hour experimental results with CaO$^+$ (blue points) and CaOH$^+$ (red points). The experimental data is averaged with different binning starting points to comply with the theoretical calculations (Appendix A). Expected bin value distribution with 0.03 probability of dark result is plotted as prediction noise. The theoretical predictions for the bin value distributions with BBR temperatures of 300\,K (blue line) and 450\,K (orange line) are plotted. The $\pm\sigma$ region represents the standard deviation of the ground rotational state population, calculated by averaging results from 100 two-hour simulations.}
    \label{fig:result}
\end{figure}

In Fig.\,\ref{fig:result}, we plot experimental results of six independent search trials, each trial contains 30,000 experiments and lasts about 20 minutes. 
To identify DPQL signals, the single shot measurements of the internal state of the calcium ion (1 for a dark result and 0 for a bright result) are binned and averaged with a moving window of size 20. 
In one of the search trials, we identify evidence of dipole-phonon interaction as shown in Fig.\,\ref{fig:result}(a).
The moving average value reaches 0.55 when the signal happens above the background noise level, which corresponds to a statistical significance of 4.1$\sigma$.
We further employed Monte Carlo generated simulation data to train a hidden Markov model (HMM) to identify DPQL signals in the experimental data and calculate the posterior probability associated each measurement in the identified signals (Appendix C). 
The HMM yields an average posterior probability of 99.8\% for this signal.

In our experiments, after CaO\(^+\) turns to CaOH\(^+\), we conducted the same experiments with CaOH\(^+\) to establish a control dataset. 
Because the closed-shell structure of CaOH\(^+\) results in  a \(\Sigma\) ground state, CaOH\(^+\) cannot have a \(\Omega\)-doublet structure. 
Additionally, compared to CaO\(^+\), the difference in secular frequencies between the IP and OP modes is only 1\,kHz and 2\,kHz, respectively. 
This difference falls within the power broadening range of the quadrupole transitions.
Therefore, since CaOH\(^+\) does not require alterations to the experimental procedure yet does not have the necessary internal structure for a dipole-phonon interaction, CaOH\(^+\) is a good candidate for control experiments.
We collected the same amount data of CaOH\(^+\) as we did for CaO\(^+\), and we plot the distributions of bin values of both data with bin size 20 in Fig.\,\ref{fig:result}(b). 
The expected bin value distribution for an experiment with a 0.03 probability of a dark result is plotted as a reference.
We find the control experiment results agree well with the noise prediction. 
On the other hand, the data from the DPQL search experiments with CaO$^+$ has a substantial increase in the probabilities of observing bins with more than six dark results relative to the background noise prediction, which is identified as evidence of DPQL signal during the experiments.

However, we also noticed that the DPQL signals have fewer appearances and a shorter lifetime than expected from theoretical predictions.
This discrepancy between the experimentally detected DPQL and the simulation-predicted DPQL rate may be attributed to factors such as modified thermal radiation at elevated temperatures as reported in a recent study~\cite{doi:10.1126/science.ado1001}.
Additionally, in the same study, the authors posited that it may even be possible that the local trap geometry affects the spectrum and polarization of the thermal radiation, which has been observed in a previous work~\cite{PhysRevLett.47.1592} and complicates theoretical predictions of the molecular state lifetimes. 
To investigate this discrepancy, we simulated the rotational state dynamics under different temperatures (Appendix A).
For a BBR temperature of 450\,K, the rate of appearance and lifetimes of the DPQL signals are closer to our observations as shown in Fig.\,\ref{fig:result}(b).
In addition, even though J-changing transitions from background gas collisions were barely observed in a recent study with CaH\(^+\)~\cite{doi:10.1126/science.ado1001}, such transitions are more likely to occur in CaO\(^+\) due to the smaller rotational energy gap, which is approximately one order of magnitude smaller than that of CaH\(^+\). 
In our simulation, the rotational state changes following a collision are modeled as a re-sampling of the thermal distribution, although the actual dynamics are likely more complex. 
For instance, the probability of larger energy changes resulting from collisions tends to decrease. 
This would extend the time CaO\(^+\) spends in the \(\Omega = 1/2\) structure due to the large fine structure energy difference, which further reduces the likelihood of populating the ground rotational state.

The signal is consistent with a successful adiabatic transfer of the excitation from the molecular ion to the motion \cite{mills2020dipole}. Our theoretical calculations show the adiabatic ramp of the trap frequency from 2$\pi\times$492 to 2$\pi\times$410~kHz at a rate of $\dot{\omega}_q=2\pi\times10$~kHz/ms will result in high fidelity state transfer when $\omega_{mol}$ is in the range  of 2$\pi\times$[420, 482]~kHz and detectable signal when $g_q>2\pi\times$400~Hz. Future experiments that change the rate and range of the sweep will be able to accurately measure these molecular constants. 

\section{\label{sec:con}Conclusion}
In conclusion, we implement coherent control and detection of the coupling between the $\Omega$-doublets of the ground rotational state of CaO$^+$ and its phonon mode in a room temperature system. 
Here, we present evidence of detection of a DPQL signal that is clearly above the noise threshold of the system and has a lower bound on the statistical significance of 4.1$\sigma$ and an average posterior probability of 99.8\%. 
Although the rate of detection and lifetime of the DPQL signals is smaller than expected, when one accounts for the local BBR temperature at the trapping region and the effect of collisions with background gases, the rate of detection and signal lifetime observed here fall within theoretical predictions. 
Additionally, control experiments were performed with CaOH$^+$, which does not have the necessary structure for the dipole-phonon interaction, and no evidence of the interaction appeared in the data.
In future experiments, the rate of occurrence of the dipole-phonon interaction may be greatly increased through the utilization of a cryogenic ion trap in which both the population and lifetime of the ground rotational state are greatly enhanced and the uncertainty from background collisions is greatly reduced. 
Additionally, it may be convenient and promising to prepare CaO$^+$ in higher rotational states and increase the coupling strength of dipole-phonon interaction for practical purposes. 

\begin{acknowledgments}
This work is supported by a collaboration between the U.S. DOE and other Agencies. This material is based upon work supported by the U.S. Department of Energy, Office of Science, National Quantum Information Science Research Centers, Quantum Systems Accelerator. Additional support is acknowledged from the Army Research Office (W911NF-21-1-0346) and the ARO Spectator Qubit MURI (W911NF-18-1- 0218).We thank Hao Wu and Michael Heaven for providing theoretical calculations of energy structures of CaO$^+$. We thank Eric Hudson and Wes Campbell for helpful discussions.  
\end{acknowledgments}

\appendix
\section{Rotational state dynamics simulation}
In our model, the population distribution of the internal states of CaO\(^+\) at temperature \( T \) follows the Boltzmann distribution. The population of $i$-th state \( p_i \) is given by: $p_i = g_i\exp\left(-\frac{E_i}{k_B T}\right)/Z$, where $g_i$ and \( E_i \) are the degeneracy and  energy of the \( i \)-th state respectively, \( k_B \) is the Boltzmann constant, \( Z \) is the partition function: $Z = \sum_{j=1}^{\infty} g_j \exp\left(-\frac{E_j}{k_B T}\right)$.
For CaO\(^+\), the vibrational constant \( \omega_e \) is 634 cm\(^{-1}\), corresponding to an energy splitting of approximately $2\pi\times$19 THz between the ground and first vibrational states. The spin-orbit constant is 130 cm\(^{-1}\), corresponding to an energy splitting of approximately $2\pi\times$4 THz between the two fine structure manifolds \( \Omega = 1/2 \) and \( \Omega = 3/2 \). This results in a 95\% population in the ground vibrational state, with the remaining 5\% in the first vibrational state, which decays at a rate of $\sim$5 Hz according to the calculation.
Within the ground vibrational state, 35\% population is in \( \Omega = 1/2 \) and 65\% in \( \Omega = 3/2 \).
In the \( \Omega = 3/2 \) manifold, the rotational levels are spaced according to \( 2(J+1)B_e \), where the rotational constant \( B_e \) is 0.37\,cm$^{-1}$ ($2\pi\times$11 GHz). This results in a broad population distribution over the rotational states. In particular, the popultion of the ground rotational state $|\Omega = 3/2, J=3/2\rangle$ is 0.47\% under room temperature $T=300$\,K.  

Both the vibrational and rotational populations are continuously perturbed by blackbody radiation 
 (BBR) from the environment, except the transitions between the two fine structure states since they are electric-dipole-forbidden. In our model, we assume an ideal blackbody environment, with an energy density $ \rho_{\text{BBR}}(\nu, T) = \frac{8\pi h \nu^3}{c^3} \frac{1}{\exp(h\nu/k_B T)-1} $ described by Planck’s law, where $\nu$ is the frequency of the radiation and $h$ is the Planck constant. The dynamics of the populations $N_{v,j}$ follow the rate equation:
\begin{widetext}
\begin{equation}
\begin{split}
\frac{dN_{v,j}}{dt} = &\sum_{E(v',j')>E(v,j)}^{v_{max},j_{max}} A_{(v',j')\rightarrow(v,j)} N_{v',j'} - \sum^{E(v',j')<E(v,j)}_{v_{0},j_{0}} A_{(v,j)\rightarrow(v',j')} N_{v,j} \\ 
&+ \sum_{v'=0,j'=0,(v',j')\neq (v,j)}^{v_{max},j_{max}}  \rho_{\text{BBR}}(|E_{v',j'}-E_{v,j}|, T) B_{(v',j')\rightarrow (v,j)}(N_{v',j'} - \frac{g_{j'}}{g_{j}}N_{v,j}),
\end{split}
\end{equation}
\end{widetext}
where $A$ and $B$ are the corresponding Einstein coefficients. 

\begin{figure}[h]
    \centering
    \includegraphics[width=\linewidth]{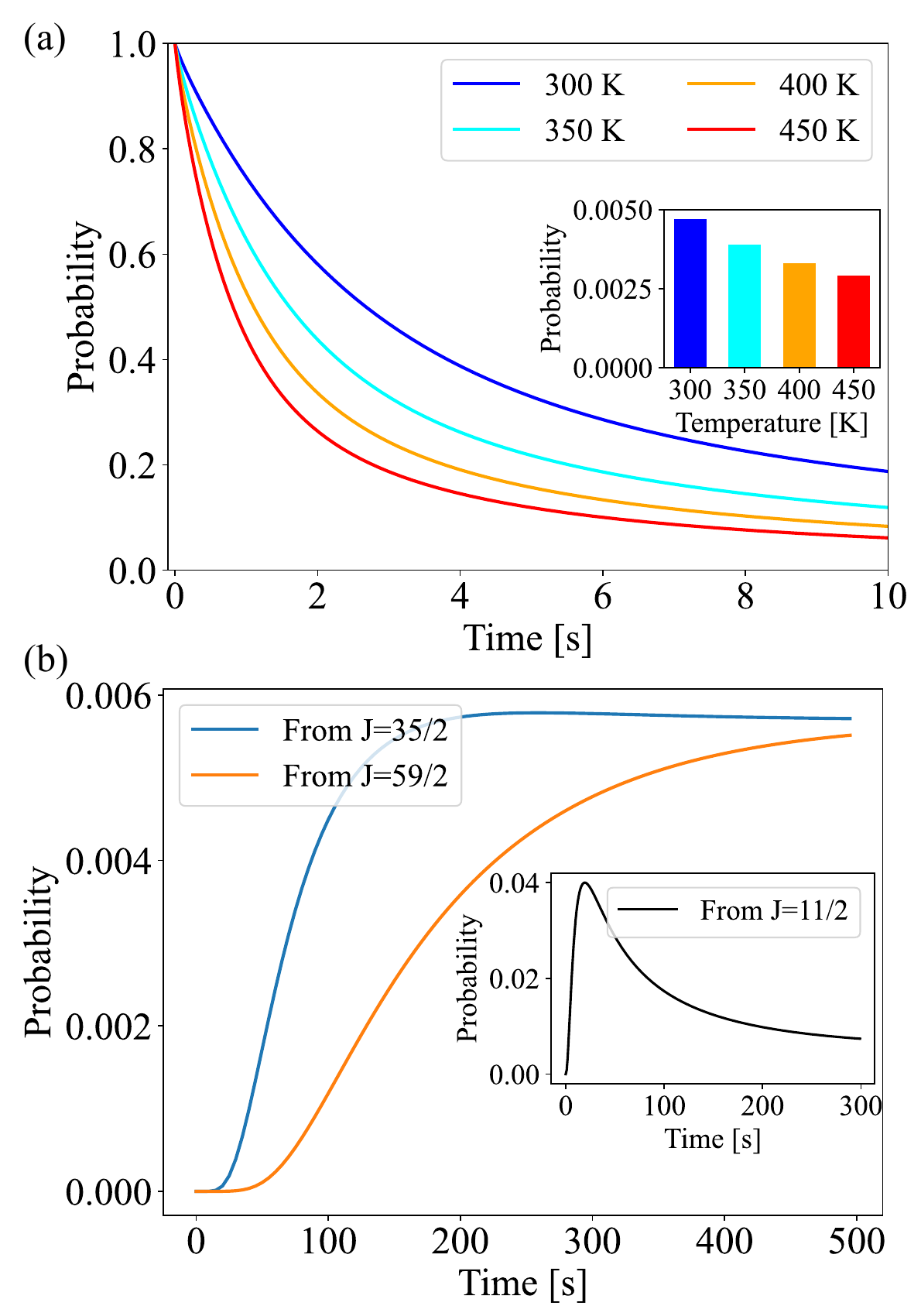}
    \caption{Simulation of rotational state dynamics of CaO$^+$. (a) Lifetime of ground rotational state under different BBR temperatures. (Inset) Thermalized population of ground state under different BBR temperatures. (b) Population evolution of ground rotational state with initial population at J=35/2 (most probable state at room temperature), J=59/2 and J=11/2 (Inset). }
    \label{fig:simulaitonlifetime}
\end{figure}

At room temperature (300\,K) and higher, the energy density around \( \omega_e \) is approximately three orders of magnitude greater than that between rotational states. Even though the dipole moment between adjacent vibrational states is about one order of magnitude smaller than between rotational states within the same vibrational state, transitions between vibrational states dominate the process. Therefore, we calculate evolutions of 70 rotational states in the ground and 70 rotational states in the first vibrational manifolds within the \( \Omega = 3/2 \) fine structure in our simulation. The population evolution of the ground rotational state is shown in Fig. \ref{fig:simulaitonlifetime}. It takes approximately 200\,s for the ground rotational state to re-thermalize from the most populated state \( J = 35/2 \) and even longer for higher rotational states such as  \( J = 59/2 \). If the initial population is close to the ground rotational state, the ground state population can reach higher value transiently due to the selection rule. Considering the simulation results, we choose 300\,s as the thermalization time of CaO$^+$.

\begin{figure}[h]
    \centering
    \includegraphics[width=0.75\linewidth]{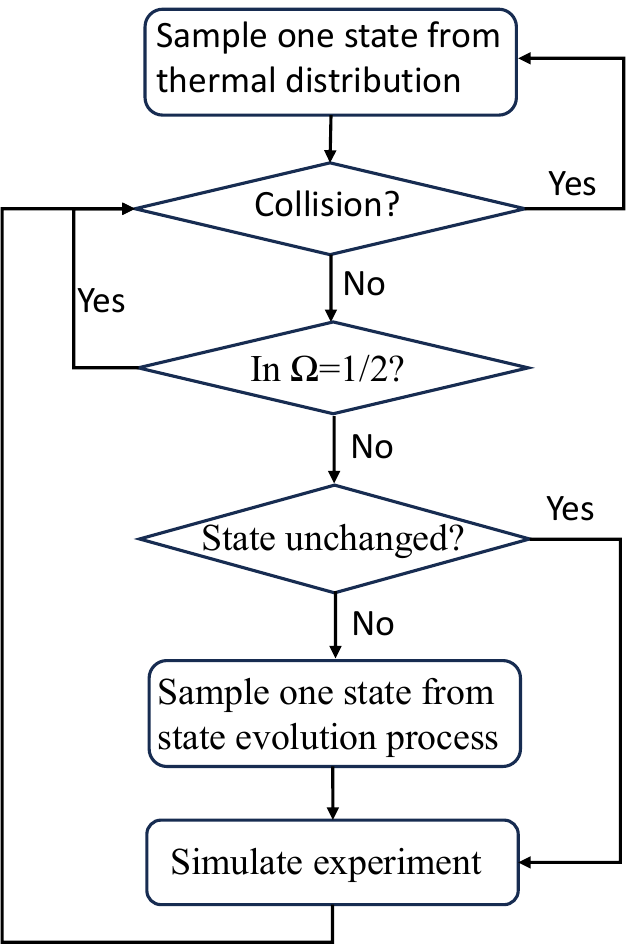}
    \caption{Flow chart of the Monte-Carlo simulation program that is used to track the rotational states of single CaO$^+$. The "Simulate experiment" part doesn't change the rotational state.}
    \label{fig:flowchart}
\end{figure}

In our model, the transfer between two fine structure states occurs via collisions with background molecules, after which the population statistically resamples the Boltzmann distribution at temperature \( T \). The collision rate is determined by the position exchange rate of the two ions, which in our system is 0.008\,s\(^{-1}\). The corresponding vacuum pressure of 1\(\times 10^{-10}\)\,Torr can be inferred from this rate assuming Langevin collisions\cite{OKA2021111482}, which is consistent with our expectations. We use a Monte Carlo method to track the rotational state for each experiment. The flow chart of the simulation is shown in Fig.~\ref{fig:flowchart}.

\section{Background noise analysis}
\begin{figure}[h]
    \centering
    \includegraphics[width=\linewidth]{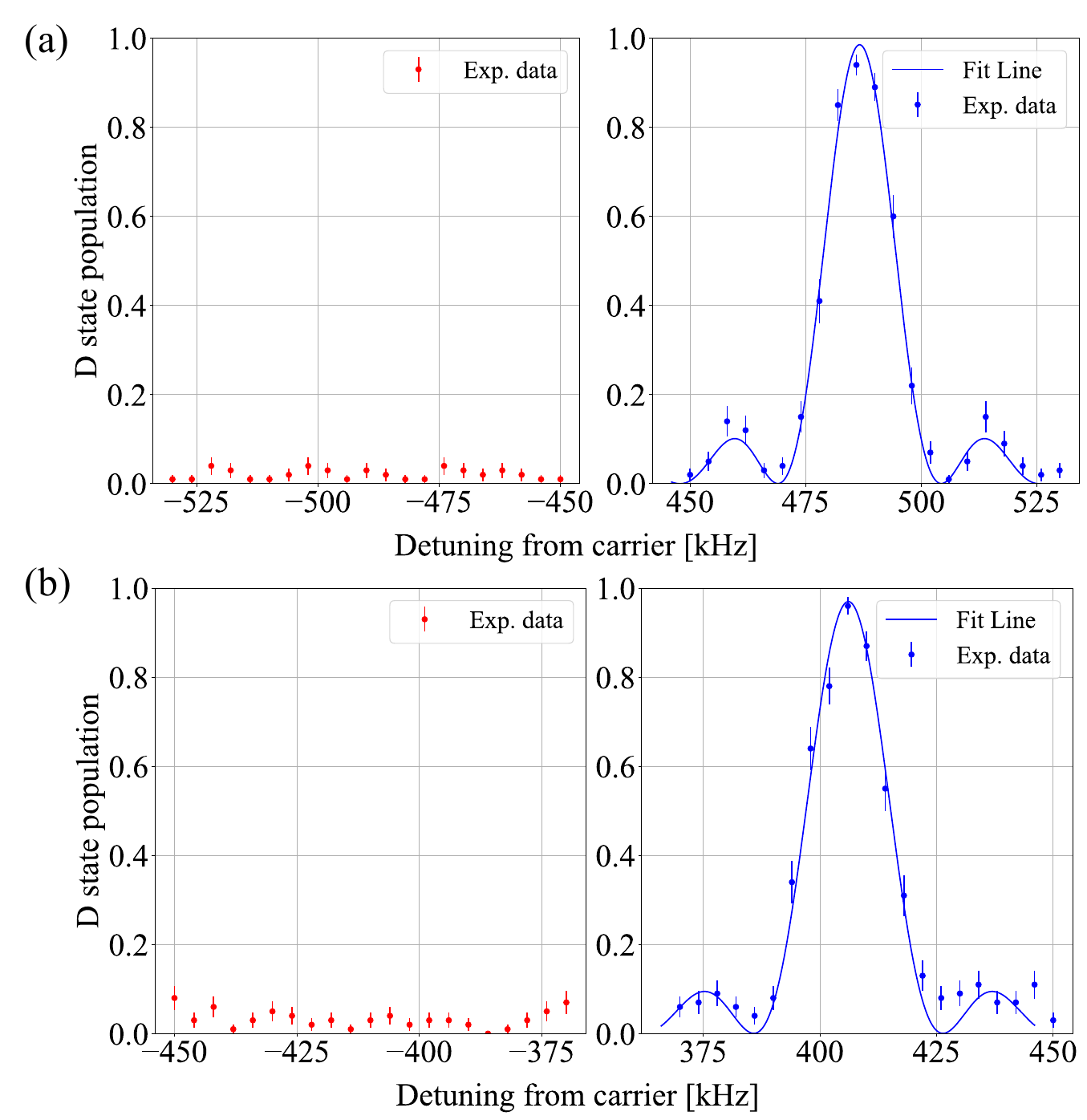}
    \caption{Sideband scan data for the OP mode (a)before and (b)after ramping the frequency from $2\pi\times$492~kHz to $2\pi\times$410~kHz. Each point is an average of 100 experiments. }
    \label{fig:rampingdata}
\end{figure}

In our experiment, the background noise, which is the false positive results, is due to Ca\(^+\) being shelved in the \(D_{5/2}\) state during red sideband operations, even without dipole-phonon interaction. With a noise level of 0.03, we attribute this mainly to off-resonant coupling to the carrier transition. Fig.~\ref{fig:rampingdata} shows red and blue sideband scan data before and after adiabatic ramping of the axial potential. The red sideband signal remains around 0.03 in both cases, indicating minimal motional state change after ramping. The blue sideband excitation fidelity is approximately 0.95 in both plots. We performed a purification procedure, similar to that in Ref\cite{Chou2017}. This process clears $|n=1\rangle$ population before detecting the red sideband signal. In this process, we find the red sideband excitation rate remained 0.03 compared to $<0.01$ theoretical noise level. We measured an environmental magnetic field noise of 9 mGauss, which causes a 9 kHz oscillation in the carrier frequency. With a carrier Rabi frequency of \(2\pi \times 90\) kHz and a detection frequency detuned by \(2\pi \times 410\) kHz with a 45\,$\mu$s pulse length, this oscillation induces an off-resonant amplitude of up to 0.044, which aligns with our experimental observations.

\begin{figure}[h]
    \centering
    \includegraphics[width=0.9\linewidth]{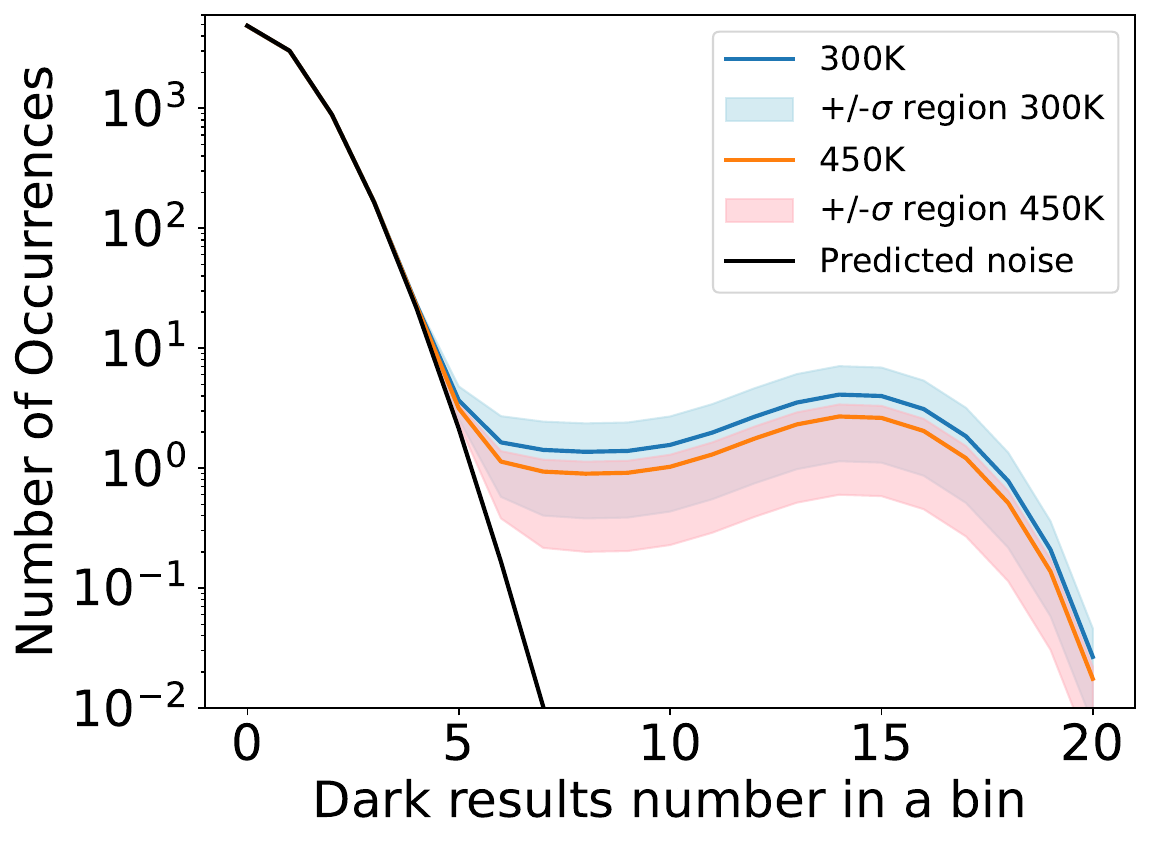}
    \caption{Predictions of Dark results number distribution of two-hour experiment data for control experiment with CaOH$^+$ and experiment with CaO$^+$at temperature of 300\,K and 450\,K. The uncertainty of ground rotational population accounts for the filled region. The predicted noise is calculated from a binomial model with binomial probability 0.03. }
    \label{fig:darkresultdistribution}
\end{figure}

In the experiment, we use a bin size of 20 to average our signals. This bin size is chosen so that the measurement uncertainty is small enough to distinguish background and DPQL signal, while the rotational state remains in the same bin. For a background noise $p_b=$0.03, the possibility of seeing $k$ false positive signals is $p_n(k)={20\choose k}p_b^k(1-p_b)^k$. On the other hand, assuming the molecular ion is in ground rotational state at the beginning of one bin, the possibility of $k$ dark results in the bin is: 
\begin{eqnarray}
p_e(k)&&=\sum_{j<k}^k \sum_{i\geq j}^{20} {i \choose j}p_d^j(1-p_d)^{i-j}{20-i \choose k-j} p_b^{k-j} \nonumber \\
&&\times (1-p_b)^{20-i-k+j} (1-p_s)^{i-1} p_s,
\end{eqnarray}
where $p_d=0.72$ is the fidelity of experiment and $p_s$ is the probability of leaving out of ground rotational state within one experiment cycle (40\,ms) at room temperature. It is 0.015 for $T=300$\,K and 0.035 for $T=450$\,K. 
For a search trial containing N experiments, the dark results distribution is calculated as:
\begin{equation}
P_t(k)=N\times[(1-p_g(T))p_n(k) + p_g(T) p_e(k)],
\end{equation}
where $p_g$ is the ground state population under temperature $T$.
For experiment length of 2 hours, we use Monte Carlo method method to simulate $p_g(T)$. Average over 100 data points, $p_g(300)=0.42\%\pm0.30\%$ and $p_g(450)=0.20\%\pm0.14\%$ compared to theoretical value 0.47$\%$ for 300\,K and 0.28$\%$ for 450\,K.
With these $p_g$ value, we calculate the bin value distribution for both with and without dipole-phonon interaction and plot them in Fig. \ref{fig:darkresultdistribution}. The one without dipole-phonon interaction (predicted noise) is used to compare with the noise level of the control experiment, which is running with molecular ion CaOH$^+$. The others are used as predictions of expected signal of experiments with CaO$^+$.

\section{Analysis of experimental data}
Here, we demonstrate how we calculate the statistical significance of a potential DPQL signal, which we present as evidence in our case for observation of the dipole-phonon interaction.
First, we calculate the probability that the suspected signal could arise from random fluctuations in background noise. 
This is quantified by a metric known as a $p$-value.
Typically for a $p$-value test, which is commonly used in biology and chemistry, a signal is accepted if the $p$-value takes on a value less than $0.05$. 
Here, we will be more scrupulous.
If one assumes a Gaussian distribution of noise in the measurement, then the suspected signal is said to have statistical significance $Z\sigma$, where $Z$ is the number of standard deviations outside of the expected noise distribution. 
This formalism is more common in the physical sciences than the $p$-value test \cite{Aad_2012_Higgs, Abbott_2016_LIGO}, and the $p$-value can be converted to this notation. 
In the $Z\sigma$ formalism, it is generally accepted that $Z\geq3$ is evidence for a signal and $Z\geq5$ is observation of a signal \cite{Lista_2016}. 
In addition to the $p$-value test, we calculate the posterior probability for the suspected signal, which is the probability that the signal can be attributed to a DPQL event given the known prior probabilities. 
To calculate the posterior probability for a potential signal, we process the data through a hidden Markov model (HMM) trained on simulated data generated using Monte Carlo methods. 

\subsection{Calculation of statistical significance: $p$-value and $Z\sigma$}
\label{suppmat:sec:StatSig}
\begin{figure}[t]
    \centering
    \includegraphics[width=\linewidth]{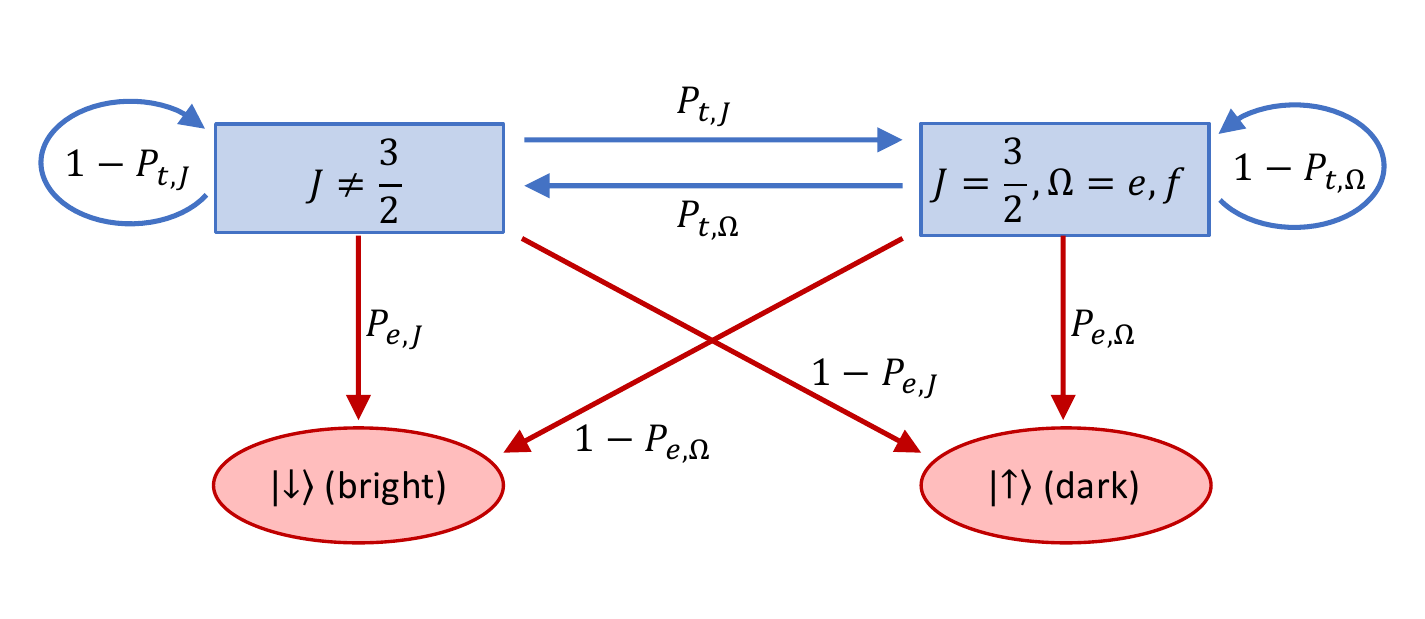}
    \caption{Diagram of the hidden Markov model used to analyze experimental DPQL data. 
    Here, the hidden states are the rotational states of the CaO$^+$, which are in blue: $J\neq 3/2$ or $J=3/2$ in which case the $\Omega$-doublet is populated. 
    The blue arrows represent the probability of transferring from one hidden state to the other (transmission probabilities, $P_t$).
    These are found via the probability distribution of rotational states and the lifetime of $J=3/2$.
    The observable states, in red, are the internal states of the Ca$^+$: bright ($S_{1/2}$, or spin down) and dark ($D_{5/2}$, or spin up). 
    The red arrows are the emission probabilities, $P_e$, or the probability that a certain hidden state will result in a certain observable state. 
    These are determined by the expected fidelity of DPQL and the readout probabilities, which are affected by noise and decoherence.}
    \label{fig:HMM}
\end{figure}

Consider the two rotational states within CaO$^+$: $J\neq 3/2$ and $J = 3/2$.
The probability that $J\neq 3/2$ produces a dark state in the Ca$^+$ should be $1-P_{e,J}=0$ but the measurement is limited by background photon counts in the experimental apparatus.
In typical experiments, this value rises to approximately $0.03$.
The probability that $J = 3/2$ produces a dark state in the Ca$^+$ should be $P_{e,\Omega}=1$ but is primarily limited by the fidelities of adiabatic ramping operations and atom shelving operation.
A conservative estimate of the fidelity of these operations is $0.72$.
Here, $P_{e,J}$ and $P_{e,\Omega}$ are known as emission probabilities in the context of HMMs and will be discussed further in the next subsection (see Fig.~\ref{fig:HMM}).
Therefore, because we expect the CaO$^+$ to be in the state $J\neq 3/2$ most of the time, in a typical experiment the result is expected to be mostly bright measurements of the Ca$^+$. 
However, if the CaO$^+$ is found to be in the $J = 3/2$ state, then we expect a string of consecutive dark measurements before the rotational state rethermalizes to the background BBR. 

Now, consider that out of $n$ total DPQL experiments (e.g., 1000 trials), there is a potential signal that consists of $x$ (e.g., 10) consecutive dark measurements. 
The $p$-value for this potential signal of length $x$ is the probability that these consecutive measurements would occur if the rotational state was $J\neq 3/2$ due to the noise in the experiment.
This probability can be calculated from the cumulative probability function, which describes the probability that the length of a string of consecutive dark measurements will take on a value equal to or less than $x$ \cite{Schilling_longestrunofheads}:  
\begin{equation}
\label{eq:cumProbFunc}
    F_{n, p_J}(x)=\sum_{k=0}^n C_n^k(x) P_{e,J}^{n-k}(1-P_{e,J})^k 
\end{equation}
where $1-P_{e,J}=0.03$ is the probability of measuring a dark state given $J\neq 3/2$ and 
\begin{equation}
    \label{eq:nChoosek}
    C_n^k(x)=\sum_{j=0}^x C_{n-1-j}^{k-j}(x) \; .
\end{equation}
\begin{figure}[t]
    \centering
    \includegraphics[width=0.8\linewidth]{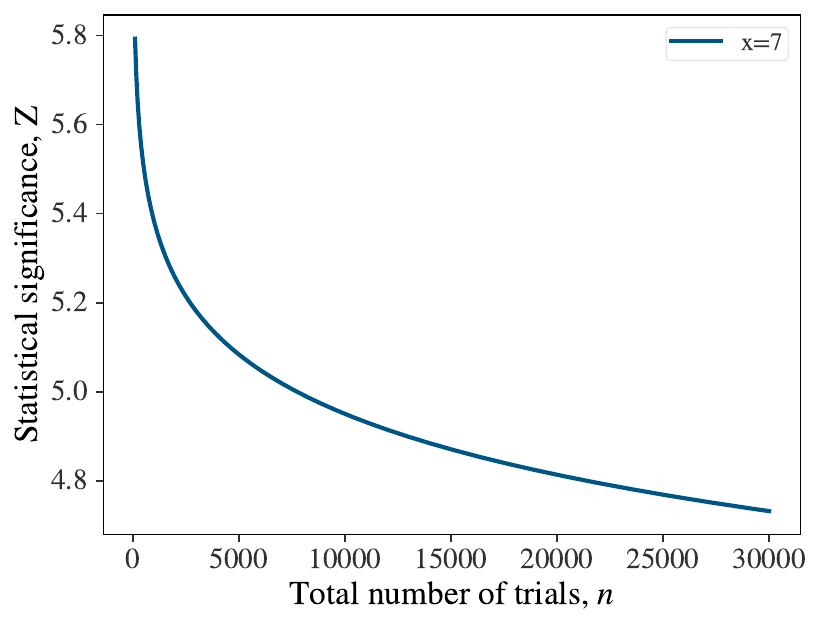}
    \caption{The statistical significance, $Z\sigma$, as a function of the total number of trials (or, experiments) for a given signal length of $x=7$.
    Because the relationship between the $p$-value for a given signal and the number of total trials $n$ is linear, we can therefore extrapolate the $p$-value to large $n$ beyond what is directly calculable. 
    Then, we can convert the extrapolation to $Z\sigma$ notation via Eq.~\ref{eq:Zsigma}.
    }
    \label{fig:statSig_Z}
\end{figure}
Using these equations, we can find the probability that random fluctuations in the DPQL experiment would allow for consecutive dark state measurements of length at least as long as $x$ given the rotational state $J\neq 3/2$: $p = 1-F_{n, P_{e,J}}(x)$. 
This is the $p$-value for the suspected signal. 
The $p$-value can be converted to the $Z\sigma$ notation via the relation \cite{Lista_2016}
\begin{equation}
    \label{eq:Zsigma}
    Z=\sqrt{2} \operatorname{erf}^{-1}(1-2 p) \; .
\end{equation}
However, for a signal of length $x$, the statistical significance changes as a function of the number of experimental trials $n$.
For example, if $n=1000$, we find that $x=4$ gives evidence for DPQL ($Z \geq 3$) while $x=7$ demonstrates observation of DPQL ($Z \geq 5$), but this would not be the case for $n=30,000$. 
See Fig.~\ref{fig:statSig_Z}.
Additionally, the recursive nature of Eq.~\ref{eq:nChoosek} makes this method of statistical analysis very computationally expensive, and the case of $n=1000$ is close to the limit of the computational ability of a personal computer. 
Nevertheless, because of the linear relationship between the $p$-value for a given signal length $x$ and the number of trials $n$, this limitation can be circumvented via numerical extrapolation as shown in Fig.~\ref{fig:statSig_Z}. 
Moreover, due to noise in the data (e.g., background photon counts, decoherence, etc.), the measured string of consecutive dark state measurements of length $x$ is likely much shorter than the true signal.
Therefore, the above calculation is treated as a lower bound on the statistical significance.

\subsection{Posterior probability and the hidden Markov model}
\label{suppmat:sec:HMM}
\begin{figure}[t]
    \centering
    \includegraphics[width=0.85\linewidth]{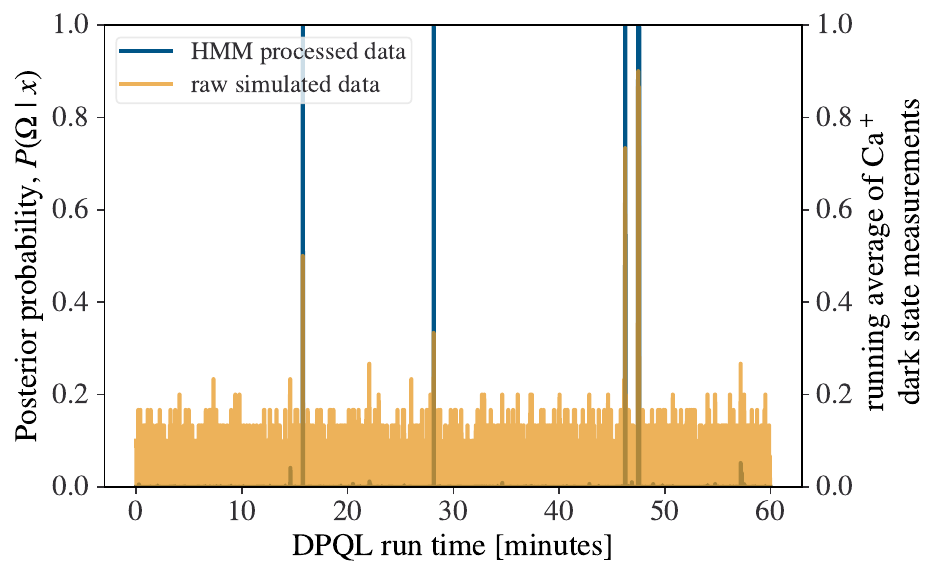}
    \caption{Simulated DPQL data produced by Monte Carlo simulation (yellow, right y-axis) and process by the hidden Markov model (navy, left y-axis).
    This demonstrates how the HMM removes noise from the data and assigns posterior probabilities to each signal.
    }
    \label{fig:simdata_HMMprocessed}
\end{figure}

While the calculation of the $p$-value and $Z\sigma$ rules out the possibility that the observed signal could be due to fluctuations in noise, the calculation of posterior probability determines the probability that the observed signal is due to the dipole-phonon interaction based on known prior probabilities. 
Along with Eqs. \ref{eq:cumProbFunc} and \ref{eq:nChoosek}, one could calculate the posterior probability using Bayes' theorem \cite{Lista_2016}:
\begin{equation}
    \label{eq:ch7:BayesTheorem}
    P(\Omega | x)=\frac{P(x | \Omega) P(\Omega)}{P(x)}=\frac{P(x | \Omega) P(\Omega)}{P(x | \Omega) P(\Omega)+P(x | J) P(J)} \; .
\end{equation}
However, to process the large datasets with little computational expense, we utilize a hidden Markov model, which generates as an output of the model posterior probabilities for potential DPQL signals. 
In a hidden Markov chain, there are observable states, the internal state of the Ca$^+$ ion (bright or dark), and hidden states, the rotational state of the CaO$^+$ ($J\neq 3/2$ and $J = 3/2$).
Additionally, to represent the dynamics of the system, we define the probability of a certain hidden state producing given observable state, the emission probabilities ($P_e$), and the probability of a certain hidden state transferring to another hidden state, the transmission probabilities ($P_t$). 
The emission probabilities, $P_{e,J}$ and $P_{e,\Omega}$, are primarily limited by noise and decoherence in the experiment and were given in the above subsection.
Meanwhile, the transmission probabilities are governed by the rotational state dynamics and depend on the temperature of the ambient BBR. 
A diagram of the hidden Markov chain is shown in Fig.~\ref{fig:HMM}.
For our analysis, we trained an HMM on 1000 hours of simulated DPQL experiments produced via Monte Carlo techniques, and we used this model to both remove noise from the experimental data and assign new posterior probabilities to the HMM-processed DPQL signals.
To enable this means of data processing, we utilized the hmmlearn package for python\cite{hmmlearn}.
The model takes in a list of observable states and outputs a prediction of the corresponding hidden state that generated a given observable state along with an associated posterior probability for each prediction based on the defined transmission and emission probabilities. 
Fig.~\ref{fig:simdata_HMMprocessed} shows both one hour of simulated DPQL trials and the posterior probability for the simulated data as provided by the HMM.

The performance of an HMM can be quantified by three machine learning metrics: precision, recall, and the F1 score. 
We can calculate these quantities by comparing the predictions of the HMM to the output of the Monte Carlo simulation before noise is injected (i.e., the true hidden state).
Precision quantifies the number of times the model incorrectly predicted a positive event ($J=3/2$) when the true state was negative ($J\neq3/2$).
The precision is calculated as
\begin{equation}
    \text{precision} = \frac{CP}{CP+IP} \; ,
\end{equation}
where $CP$ is the number of times the model correctly predicted a positive event and $IP$ is the number of incorrectly predicted positives (negatives identified as positive). 
Our model achieved a precision score of $0.98$.
Recall quantifies the number of times the model missed a positive event by predicting a negative.
The recall is calculated as 
\begin{equation}
    \text{recall} = \frac{CP}{CP+IN} \; ,
\end{equation}
where $IN$ is the number of incorrectly predicted negative events. 
The recall score for our model is $0.97$.
Finally, the F1 score is a combination of precision and recall into a single metric:
\begin{equation}
    \text{F1} = \frac{\text{precision}\cdot \text{recall}}{\text{precision}+\text{recall}}
\end{equation}
The F1 score for our model is $0.97$.
The above scores represent the performance of our HMM that was trained on 1000 hours of simulated DPQL data with a combined operational fidelity of 0.72. 
When models were trained with different levels of operational fidelity, it was found that a model that was trained on data with lower fidelity would still perform well on data with both low and high operational fidelity. 
However, a model that was trained on near unity fidelity could not accurately predict the hidden state from the data of observed states, achieving an F1 score as low as $0.20$.
In the analysis of the experimental data, unless otherwise stated, we use the model that was trained with an operational fidelity of 0.72 as this is based on the expected noise and motional decoherence in our experimental system.

\bibliography{DPQLarxiv}

\end{document}